\documentclass [amssymb,  amsmath, showpacs]{revtex4} 
\usepackage{bm}
\usepackage{graphicx,graphicx,epsfig}
\usepackage{color, soul}

\newcommand{\bP}{\mathbf{P}}

\newcommand{\la}{\langle}
\newcommand{\ra}{\rangle}
\newcommand{\rar}{\rightarrow}
\newcommand{\da}{\downarrow}
\newcommand{\ua}{\uparrow}
\newcommand{\be}{\begin{eqnarray}}
\newcommand{\ee}{\end{eqnarray}}

\begin{document}


\title{ Higher Order Spin Noise Statistics}
\author{ Fuxiang Li$^{1, 2,3}$, Avadh Saxena$^3$, Darryl Smith$^3$, Nikolai A. Sinitsyn$^{3,*}$}
\address{$^1$ Center for Nonlinear Studies, Los Alamos National Laboratory,  Los Alamos, NM 87545 USA}
 \address{$^2$ Department of Physics, Texas A\&M University, College Station, TX 77845 USA}
\address{$^3$ Theoretical Division, Los Alamos National Laboratory,   Los Alamos, NM 87545 USA}
\email{ nsinitsyn@lanl.gov}

\date{\today,\now}

\begin{abstract}

The optical spin noise spectroscopy  (SNS) is a minimally invasive route towards obtaining dynamical information about electrons and atomic gases by measuring mesoscopic time-dependent spin fluctuations.  Recent improvements of the sensitivity of SNS should make it possible to observe higher order spin correlators  at thermodynamic equilibrium. We develop theoretical methods to explore higher order  (3rd and 4th) cumulants of the spin noise  in the frequency domain. We make predictions for the possible functional form of these correlators in single quantum dot experiments and then apply the method of the stochastic path integral to estimate effects of many-body interactions.
\end{abstract}

\date{\today}
\maketitle

\medskip



\section{\label{sec:level1} Introduction}

The full counting statistics has become a widely discussed topic both in electronics and quantum optics \cite{fcs1,fcs2,fcs3,fcs4,fcs5, fcs6}. Its measurements promise to provide considerably more information about interacting electrons and photons than that could possibly be obtained from standard  linear response characteristics.

\begin{figure}
\includegraphics[scale=0.5]{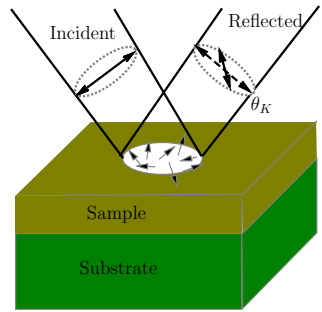}
\caption{\label{fig:schematic} A sample with quantum dots or conducting electrons in a semiconductor is illuminated by a polarized laser beam, and the Faraday or Kerr rotation angle of its polarization plane $\theta_K$ is measured. }
\end{figure}

Spins in a small mesoscopic volume of a semiconductor with $N$ electrons  typically experience statistical fluctuations of order $\sqrt{N}$, even in a zero magnetic field at the thermodynamic equilibrium.  The {\it spin noise spectroscopy} (SNS) is an optical technique that provides a viable route to study such fluctuations by directly measuring local spin correlators \cite{Aleksandrov81, Crooker04}. Experimentally, spin correlators were obtained by measuring the rotation of the linear polarization of a light beam that passes through a mesoscopic region with spins, e.g. as shown in Fig.~\ref{fig:schematic}. The polarization rotation angle is proportional to the local instantaneous magnetization of the region and can be traced with sub-nanosecond resolution in time \cite{Minhaila06_2}. Characterization of the equilibrium spin noise has been demonstrated and used to determine g-factor, spin coherence and spin relaxation times of electrons in GaAs \cite{Oestreich05, Muller08, Crooker09, Oestreich10, cond-SNS, Huang11, ultrahigh} and atomic gases \cite{Crooker04, Minhaila06, Minhaila06_2, Katsoprinakis07, Shah10, Zapasskii13}. An application of SNS to the central spin dynamics in InGaAs hole doped quantum dots \cite{Crooker10, Li12, WarburtonArxiv13,roy-13}, in particular, revealed the importance of the nuclear quadrupolar coupling for the decoherence and relaxation of a spin qubit \cite{sinitsyn-12prl}.

 Higher order spin correlators are not the subject of application of the standard fluctuation-dissipation relations. Hence, by studying higher order statistics, one can obtain information about spin dynamics that simply cannot be found in the average linear response characteristics of spin systems \cite{fcs-exp1, fcs-exp2, fcs-exp3}. For example the so called phase transitions at fluctuating level   attract lots of interest  \cite{rare-pt}. Recently, it was realized that such unusual critical phenomena can be effectively studied in physical systems by measuring time-dependent behavior of high order noise cumulants \cite{fcs-exp3,flindt-13pre}.

Most of the studies, however, have been focused so far on the  spin noise power spectrum  \cite{simple-noise, Kos10, Glazov11, Pershin12, sinitsyn-12prl} and the associated  second order spin correlator
\begin{equation}
g_2(t)=\la S_z(t) S_z(0) \ra,
\label{c2-22}
\end{equation}
where $S_z(t)$ is the time-dependent spin polarization in a mesoscopic region and $z$ is the measurement axis.
The discussion  of higher order spin correlators at the thermodynamic equilibrium has been essentially absent.  This is partly due to the fact that for a large number $N$ of spins the physical noise is dominated by Gaussian fluctuations
 which are fully described by (\ref{c2-22}), as articulated by the law of large numbers.  Hence, in order to reveal the additional information about spin dynamics in the form of higher order cumulants one should not only filter the useful signal from the background noise but also filter out the physical Gaussian fluctuations. In noninteracting spin systems, non-Gaussian effects are due to discreteness of spin states.

 Among the most relevant theoretical and experimental work, which refered to the optical spin noise spectroscopy, we mention here the  model of a weak measurement framework for a single spin system, developed in \cite{sham}, which discusses time interval statistics between detector clicks. The type of time correlators studied in that work, however, is somewhat distant from the one that is currently accessible by SNS.
On the experimental side, in    \cite{fcs-exp4} and  \cite{simpleexp}, simple examples of  measurements of higher order noise statistics of acoustic sound in the frequency domain were provided, claiming that similar methodology may work to study the spin noise in an SNS setup.

 SNS appears to be particularly promising for characterization of the high order spin statistics. It provides a considerable and continuous stream of data for statistical analysis. Information can arrive with the rate of gigabites per second and the measurement time is formally unrestricted, e.g. it can be several weeks if needed. Moreover, the sensitivity of SNS has been continuously improving. For example, recently introduced Ultrahigh Bandwidth SNS achieved picosecond time-scale resolution \cite{Shah10}. New measurement schemes have  also been proposed  recently to increase the polarimetric sensitivity by using high polarization exctinction schemes and hence suppress the relative role of the background noise sources  \cite{alex}.  Noise of a single spin of a heavy hole localized in a flat (InGa)As quantum dot has been successfully observed recently by SNS   \cite{Dahbashi13}. The latter achievement is an important milestone on the way to experimental studies of the higher order spin statistics because the law of large numbers no longer applies to $N=1$, which means that a number of higher order cumulants of spin noise statistics should be of the same order of magnitude as the spin-spin correlator (\ref{c2-22}). Non-optical methods have also been developed and demonstrated recently to probe correlators of single spins \cite{STM,nV}, which may  be considered for studies of higher order spin noise statistics.

It is also helpful that optical SNS is free from many of the problems that were encountered in measurements of higher order statistics of electric currents \cite{josephson, Hinich05, Elgart04}.  In SNS, since spin fluctuations are probed optically, there is no problem with the often detrimental noise from leads, and one generally does not need a complex nano-lithography. The advancement of the spin noise spectroscopy thus motivates us to reconsider the modern technological capabilities of using  higher order noise statistics measurements as an important tool for materials studies and quantum information science.

\begin{figure}
\includegraphics[scale=0.6]{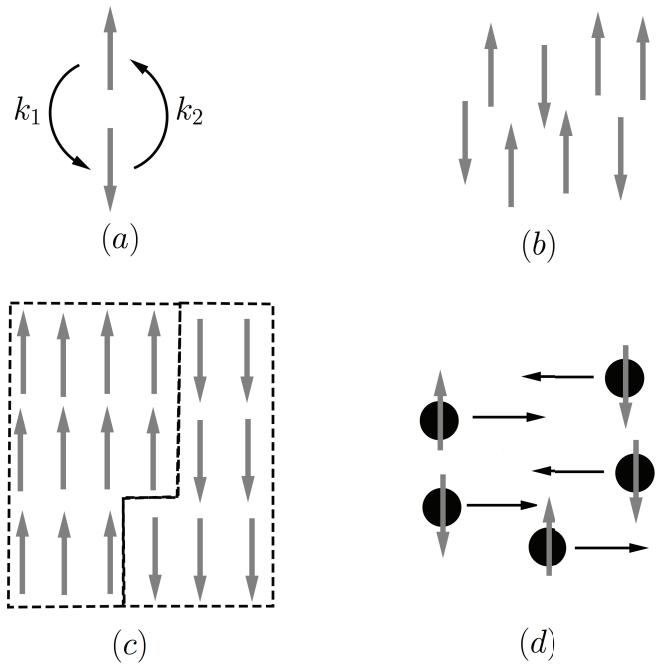}
\caption{\label{fig:model} Four different systems of our focus: (a) A single Ising spin with two states, up and down, between which transitions can happen with rates $k_1$ and $k_2$. (b) $N$ noninteracting Ising spins. (c) Model with a ferromagnetic coupling leading to long-range correlations. (d) Model of conducting electrons with Pauli exclusion interactions. }
\end{figure}

The plan of our article is as follows.  In section \ref{sec:2}, we define the 3rd and 4th order spin noise cumulants and discuss their properties.
 We will then consider four different applications (Fig.~\ref{fig:model}) in which we believe the experimental observation of higher order spin correlations is most likely:
 In section \ref{sec:single spin}, motivated by the observation of the spin noise of a single central spin in a quantum dot, we explore a possible form of 3rd and 4th order  correlators of a single spin (Fig.~\ref{fig:model}(a))  and make specific predictions for verification in InGaAs quantum dots \cite{Dahbashi13}. In section \ref{sec:simple}, based on a simple non-interacting spin model, we develop a method based on the stochastic path integral to calculate spin noise statistics and, for illustration, we apply it to an ensemble of a mesoscopic number of noninteracting spins (Fig.~\ref{fig:model}(b)). Then in section \ref{sec:ferro} we explore the effect of many-body interactions, such as the ones arising near a ferromagnetic phase transition in magnetic semiconductors (Fig.~\ref{fig:model}(c)).  We apply the path integral method to a phenomenological kinetic model based on the Glauber spin dynamics to demonstrate that higher order spin noise statistics becomes particularly insightful to observe near a phase transition point.
 In section  \ref{sec;Pauli}, we explore the  higher order spin cumulants  for conducting electrons (Fig.~\ref{fig:model}(d)) and explore the effect of the Pauli principle.

\section{\label{sec:2}Properties of 3rd and 4th order spin correlators in frequency domain}

Complete information about an interacting many-body system is contained in the full set of all cumulants of system variables.
Let us introduce a normalized spin polarization
\begin{equation}
\delta S_z(t) = S_z(t)-\la S_z\ra,
\label{deltas}
\end{equation}
and consider a random variable
\begin{equation}
a(\omega) = \int_0^{T_m} dt e^{i\omega t} \delta S_z(t),
\label{aom}
\end{equation}
where  $T_m$ is the time of measurement, which is assumed to be much larger than other physical time-scales in the system, including the characteristic relaxation time $T$ of the spin system. Note that since $S_z(t)$ is real, we have $a^*(\omega)=a(-\omega)$. For a paramagnetic system, one finds that $\la a(\omega)\ra=0$, where the average is considered over repeated measurements under  identical experimental conditions. The most accessible
physically interesting characteristic is the {\it  noise power},  which is defined as
\begin{equation}
C_2(\omega) = \langle |a(\omega)|^2 \rangle.
\label{noiseP1}
\end{equation}
Its knowledge is equivalent to the knowledge of a spin-spin correlator in the time domain via
\begin{equation}
C_2(\omega)=2T_m \int_0^{\infty} dt \cos(\omega t) \la \delta S_z(t) \delta S_z(0) \ra.
\label{c2-1}
\end{equation}

 At steady conditions, only products of $a(\omega_i)$ with $\sum_i \omega_i=0$ can be nonzero after averaging. Hence, the next nearest nontrivial correlator of $a(\omega)$ is of the 3rd order and  given by
\begin{equation}
C_3(\omega_1,\omega_2) = \langle a(\omega_1) a(\omega_2) a^* (\omega_1+\omega_2) \rangle.
\label{c3-1}
\end{equation}
A specific property of $C_3$ is that it is zero in a system with a time-reversal symmetry. There can be a variety of potentially interesting combinations of 4th powers of $a(\omega)$. The one that has the most transparent physical meaning is the so called {\it bispectrum}:
\begin{equation}
 C_4 (\omega_1,\omega_2) = \langle |a(\omega_1)|^2 |a(\omega_2)^2| \rangle -  \langle |a(\omega_1)|^2  \rangle \langle |a(\omega_2)^2| \rangle, \quad \omega_1\ne \omega_2.
 \label{c4-1}
 \end{equation}
For $\omega_1=\omega_2$ its definition is modified:
\begin{equation}
 C_4 (\omega) \equiv C_4 (\omega,\omega)= \langle | a(\omega)|^4  \rangle -  2\langle |a(\omega)|^2  \rangle^2.
\label{c4-2}
 \end{equation}
The bispectrum (\ref{c4-1}) indicates how spin noise components at two different frequencies ``talk" to each other. For example, if $C_4(\omega_1,\omega_2)$ is negative, one can conclude the presence of anti-correlations, i.e. observation of a strong signal at one frequency presumes that another frequency is likely suppressed, etc.

The form of, so called, cumulants (\ref{c3-1}), (\ref{c4-1}) and (\ref{c4-2}) is dictated by the requirement that they are zero for Gaussian fluctuations of $a(\omega)$, so that they do not duplicate the information that can be obtained from (\ref{noiseP1}).
One more important property of cumulants is their ``additivity", i.e.  independent noise sources additively contribute to their values. This is important because experimental measurements in SNS generally produce a strong background shot noise, which can be filtered out by separately measuring the pure background in a strong magnetic field applied in the direction transverse to the measurement axis. After subtracting the background contribution from the measured cumulants one obtains  the physical values of cumulants.

Another consequence of such an additivity of cumulants is  that they generally linearly increase with the measurement time $T_m$. Indeed, if $T_m$ is much larger than the relaxation time $T$, one can expect that a measurement during $T_m$ is roughly equivalent to $T_m/T$ independent measurements which additively contribute to the finite cumulant value. This property is one of the reasons of the difficulty in observing higher order statistics. For example, the dimensionless ratio $C_4 (\omega)/(C_2(\omega))^2$ should generally be proportional to
$T/T_m\ll1$. Experimentally, it is important to keep $T_m$ large to avoid nonphysical effects of a finite measurement time. On the other hand, one should keep the ratio  $T/T_m$ not too small in order to be able to filter $C_4 (\omega)$ from the Gaussian part of spin noise statistics.

As another word of caution, we would like here to point to a problem that has not appeared in previous measurements of the noise power spectrum (\ref{noiseP1}).
Most cumulants are protected only against an additive background noise, i.e. the noise that is uncorrelated from the physical signal. In SNS, fluctuations of spins are deduced from the fluctuations of an optical signal. The probability of light interaction with spins depends on the laser beam intensity, which is often slowly fluctuating.  For $C_2$ and $C_3$ such fluctuations merely renormalize their amplitude, keeping relative amplitudes at different frequencies the same. In contrast, the expression, e.g. (\ref{c4-1}), for $C_4$ is the difference of two terms that may be differently renormalized by the beam intensity fluctuations. Moreover, for a large number of noninteracting spins in the observation region, each of those terms can be much larger than their difference  in (\ref{c4-1}) so that even small instabilities in the beam intensity can lead to an admixture of Gaussian fluctuations and degrade the measurements of $C_4$. Hence $C_4$ should become most accessible when the number of observed spins is relatively small so that the dimensionless ratio, $C_4/(C_2)^2$, is as large as possible and the cumulants are averaged over a smaller span of time than the time scale of slow fluctuations of the beam intensity.
This situation can be realized in a single spin quantum dot as in \cite{Dahbashi13} or in the case of considerable spin correlations, e.g. when one effectively  observes dynamics of a small number of ferromagnetic domains.

Higher order  correlators depend on more than one frequency, so obviously, they contain additional and, possibly, considerably larger amount of information about the system than the noise power. In solid state applications,  as a proof of principle, $C_3(\omega_1,\omega_2)$ was measured as a function of frequencies only recently to describe the shot noise  of electric charge currents through an artificial nanostructure \cite{fcs-exp4}. Specially engineered nanoscale systems could allow measurements of charge current cumulants
in the time domain up to 15th order and the  study  of fundamental nonequilibrium fluctuation relations \cite{fcs-exp1,fcs-exp2,fcs-exp3} but such measurements could not be applied to materials characterization.
Therefore, we will explore the information that can be obtained about condensed matter systems by measuring higher order spin, rather than current, correlators.

\section{\label{sec:single spin} Single spin noise}

\subsection{3rd order correlator of Ising spin dynamics}
From the point of view of statistical filtering of a useful signal, the 3rd order correlator (Eq.~\ref{c3-1}) is the next in complexity after the popular 2nd order spin-spin correlator. However, the 3rd order correlator changes sign under  time reversal and hence its observation requires a specific breakdown of a time reversal symmetry. In a single spin InGaAs quantum dot, conditions for 3rd order cumulant observation can be created by applying a strong (of the order of 1 Tesla) magnetic field in the out-of-plane (i.e. parallel to the measurement z-axis) direction. At such fields, spin relaxation is dominated by interactions with phonons. At a sufficiently low temperature, Zeeman coupling in such a magnetic field becomes comparable to $k_BT_s$, where $T_s$ is the system temperature and $k_B$ is the Boltzmann constant.
Without an in-plane magnetic field component, one can disregard coherence effects and assume that the spin essentially behaves as a classical Ising spin. Spin transitions between up and down states can be then described by kinetic rates, respectively, $k_1$ and $k_2$, which satisfy the
detailed balance condition
\begin{equation}
k_1/k_2=\exp (-g_{z}B_z/k_BT_s),
\label{k2k1}
\end{equation}
where $B_z$ is the z-component of the magnetic field and $g_z$ is the corresponding g-factor.

\begin{figure}
\includegraphics[scale=0.5]{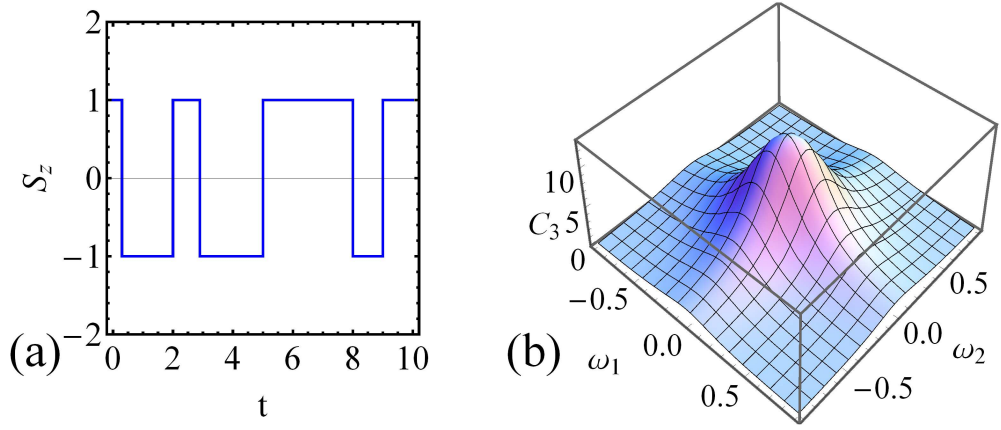}
\caption{\label{fig:C2C3}  (a) Telegraph noise induced by dynamics of a single spin polarization $S_z(t)$. (b) The 3rd order cumulant of spin noise, Eq.~(\ref{c3-2}),  for a single spin with $k_1=0.2$ and $k_2=0.1$; $T_m=1$.}
\end{figure}

Spin polarization dynamics of this system behaves like a telegraph noise, as shown in Fig.~\ref{fig:C2C3}(a). Let $p_1(t)$ and $p_2(t)$ be time-dependent probabilities of  a spin being, respectively, in the up and down states. The dynamics of probabilities is governed by the master equation:
\begin{eqnarray}
\dot{p}_1(t)&=&-k_1 p_1+k_2 p_2, \nonumber\\
\dot{p}_2(t)&=&k_1 p_1-k_2 p_2,
\label{pp12}
\end{eqnarray}
which can be solved as $\bP(t)=\mathcal{U}(t)\bP(0)$, with $\bP(t)=(p_1, p_2)^T$, where the evolution operator $\mathcal{U}$ is given by
\begin{eqnarray}
\mathcal{U}(t)=\frac{1}{k_1+k_2} \left(
                     \begin{array}{cc}
                       k_2+k_1e^{-(k_1+k_2)t} & k_2(1-e^{-(k_1+k_2)t}) \\
                       (1-e^{-(k_1+k_2)t})k_1 & k_2e^{-(k_1+k_2)t}+k_1
                     \end{array}
                   \right).
\end{eqnarray}

It can be shown that the average spin polarization is a constant:
\begin{eqnarray}
\langle S_z(t)\rangle=\langle{\bf 1}|\sigma_z\mathcal{U}(t)|\bP_0\rangle =\frac{k_2-k_1}{k_1+k_2},
\end{eqnarray}
which is equal to the average spin at $t=0$: $\la S_z\ra\equiv\langle {\bf 1}|\sigma_z|\bP_0\rangle=\frac{k_2-k_1}{k_1+k_2}$, where $\sigma_z$ is the Pauli matrix. Here $\la {\bf 1}|=(1, 1)$ and the initial spin polarization $|\bP_0\ra=(p_1, p_2)^T$ with $p_{1, 2}=k_{2, 1}/(k_1+k_2)$. Note that, for simplicity, here we normalized the Ising spin polarization values to be $\pm 1$ rather than $\pm 1/2$.

The second order correlator of spin fluctuations in real time is then given by
\begin{eqnarray}
C_2(t_1, t_2)&&=\langle \delta S_z(t_1)\delta S_z(t_2)\rangle=\frac{8k_1k_2}{(k_1+k_2)^2}e^{-(k_1+k_2)(t_1-t_2)},
\end{eqnarray}
in which $\la S_z(t_1) S_z(t_2) \ra=\langle {\bf 1}|\sigma_z\mathcal{U}(t_1-t_2)\sigma_z{\cal U}(t_2)|\bP_0\rangle$.
The spin noise power spectrum, defined as in Eq.~(\ref{c2-1}), reads:
\be
C_2(\omega)=\frac{8k_1k_2}{k_1+k_2}\frac{T_m}{\omega^2+\gamma^2}.
\label{eq:C2single}
\ee
with the relaxation rate $\gamma=k_1+k_2$.

The 3rd order  spin correlator can  be calculated similarly. For $t_1>t_2>t_3$, we have:
\begin{eqnarray}
C_3(t_1, t_2, t_3)&&=\langle \delta S_z(t_1)\delta S_z(t_2)\delta S_z(t_3)\rangle \nonumber\\
&&=\frac{8k_1k_2(k_1-k_2)}{\gamma^3} e^{-\gamma (t_1-t_3)},
\end{eqnarray}
with $\la S_z(t_1) S_z(t_2) S_z(t_3) \ra=\langle {\bf 1}|\sigma_z \mathcal{U}(t_1-t_2) \sigma_z \mathcal{U}(t_2-t_3) \sigma_z {\cal U}(t_3)|\bP_0\rangle$.

For the case of $t_2>t_1>t_3$, the expression for $C_3$ can be obtained by exchanging $t_1$ and $t_2$. Finally, by taking the Fourier transform, the 3rd order cumulant in the frequency domain is found:
\be
C_3(\omega_1, \omega_2)=T_m \frac{16 k_1 k_2 (k_1-k_2)}{k_1+k_2} \frac{\omega_1^2 +\omega_1\omega_2 +\omega_2^2 +3\gamma^2} {(\omega_1^2 +\gamma^2) (\omega_2^2 +\gamma^2) ((\omega_1^2 +\omega_2^2)^2 +\gamma^2)}.
\label{c3-2}
\ee

One can conclude from (\ref{c3-2}) that $C_3$ is nonzero only when $k_1\ne k_2$, i.e. when a strong magnetic field changes the relative rates of spin transitions. If $k_1-k_2$ changes sign, so does $C_3$. Measurements of $C_3$ in a quantum dot system can be used, e.g. as  an independent probe of the g-factor via  (\ref{k2k1}) at high values of the magnetic field. In Fig. \ref{fig:C2C3} we  plot $C_3(\omega_1, \omega_2)$ for the case of $k_1-k_2>0$.

Finally, we note that noise in simple kinetic models have been previously studied in different contexts, e.g. in statistics of electric currents coupled with a fluctuating two level degree of freedom \cite{fcs-exp1,Emary07, josephson}.
For example, spin correlator (\ref{c3-2})  has the same functional form as the third order electric current correlator derived in \cite{Emary07}.

\subsection{4th order cumulant of the Ising spin}

The advantage of the 4th order cumulant is that it is nonzero even without time symmetry breaking. For a single spin, it is expected to be nonzero because the discreteness of spin states leads to a binary signal, which is  similar to a telegraph noise rather than Gaussian fluctuations. Since strong out-of-plane magnetic fields are not needed for its observation, one can measure  $C_4(\omega)$ in a standard setting, e.g. in a zero magnetic field. The spin of a heavy hole in an InGaAs quantum dot at such conditions behaves essentially as an Ising spin with Markovian transitions between up and down states due to the coupling to a quickly fluctuating nuclear spin bath \cite{sinitsyn-12prl}.

While, in principle, analytical calculations  of $C_4$ via finding its value in the time domain are possible, we found them very tedious even for the simplest models. In the next section, we will describe an alternative approach, which is based on the stochastic path integral technique, and which provides a simpler framework for the calculation of $C_4$, including for independent Ising spins.
 In this subsection, instead, we will present  a numerical approach that simulates the weak measurement setup. Such simulations are close in spirit to the real physics encountered in the experimental measurement process of SNS and allow us, in particular, to study the effect of detector parameters on the observable higher cumulants.

As in the previous subsection, we  consider an Ising spin but assume that spin flips between up and down states happen with the same rates $k=1/T$, where $T$ is the characteristic life-time that for a hole doped InGaAs quantum dot was estimated to be $\sim 0.4\mu$s \cite{Li12}.
We simulate the weak measurement process with the scheme that was proposed for SNS in \cite{sham}. According to it, measurements are performed in discrete time steps $\tau \ll T$. Suppose that with a probability $p_D<1$ the coupling to the detector induces  the collapse of the state vector
at each time step, leading to a detector ``click", i.e. the measurement of the spin state. Respectively, with a probability
\begin{equation}
1-p_D\equiv e^{-\tau/T_D},
\label{def_td}
\end{equation}
we observe no useful signal at the detector per single measurement. In (\ref{def_td}), we introduced the time $T_D$ that characterizes the typical time between collapses of the spin vector to the measurement basis.

\begin{figure}
\scalebox{0.4}[0.4]{\includegraphics{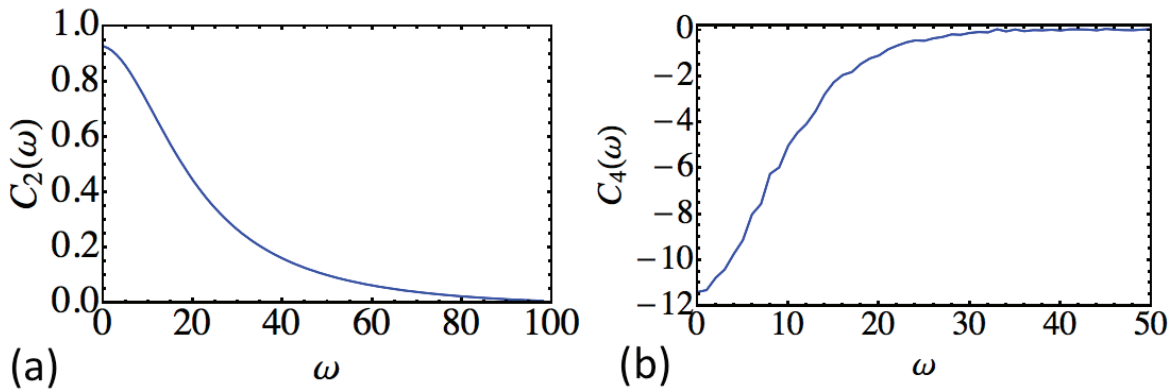}}
\hspace{-2mm}\vspace{-4mm}
\caption{ Numerically calculated (a) noise power, Eq. (\ref{noiseP1}), and (b) the 4th order cumulant, Eq.~(\ref{c4-2}), for a single Ising spin noise. Elementary measurement time step is normalized to $\tau=1$. Cumulants are normalized by dividing them by the total measurement time $T_m=256\tau$;  relaxation time  is $T=5\tau$ and $T_D=2\tau$.  Background level was subtracted by normalizing correlators at large frequencies to zero value. Averaging is over 100 million of statistically independent measurements of $a(\omega)$.
} \label{spectr}
\end{figure}
  Our measurement axis is the $z$-axis, and there are  three possible measurement outcomes. Let us call them $1,0,-1$, where $\pm 1$ correspond to the collapse of the state vector to one of the spin states $|\pm \rangle$, which also correspond to specific $\pm 1$ outputs of the detector, while $0$ corresponds to the absence of a useful detector click.

If the probability per single measurement step to collapse the spin state  is  small, most of the measurement outcomes would be zeros. However, when the collapse of the spin vector happens, we  determine the state of the spin, and the   measurement produces $+1$ or $-1$. After this, the evolution of the spin is again calculated according to the master equation.
A numerical simulation of such a process returns a random sequence that simulates the output of a detector, such as
$$
0000\,{\rm 1}\, 000000000000000\,{\rm -1}\, 000000000000\, {\rm 1}\,000\ldots, \,\,\, {\rm etc}.
$$
Evaluating it for a  span of time $T_m \sim 256\tau $, we found its Fourier transform and calculated the powers of $a(\omega)$. By repeating this process, we produced a 100 million of such random sequences, from which we
calculated  the cumulants.  As in SNS experiments, the fact that most of the time a weak measurement fails to produce the detector response  leads to a broad background white noise, which we subtracted to obtain the physical part of the spectrum.
Our results for the noise power, Eq. (\ref{noiseP1}), and for the 4th order cumulant, Eq. (\ref{c4-2}), are shown in Fig.~\ref{spectr}.
As expected, the noise power spectrum (Fig.~\ref{spectr}(a)) is Lorentzian. The 4th correlator appears to be negative and narrower than $C_2$. The ratio of the integrated cumulants
\be
\eta_1=\frac{\int d\,\omega C_4(\omega)}{ \int d\, \omega C_2(\omega)^2}
\label{eta-21}
\ee
is found to be $\eta_1=-1.9(T/T_m)$, which is close to the theoretically predicted value $\eta_1 = -2(T/T_m)$ (section IV) in the limit $T_m \gg T \gg \tau$.

The numerical simulations of measurements of $C_4$, for a single spin, converge relatively quickly when the time $T_D$ is smaller than $T$, i.e. several nonzero detector clicks happen before a spin flip.
In the opposite case, $T_D<T$, convergence of the simulations to the expected values of $C_4(\omega)$ quickly deteriorates, and the result appears to be dependent on the measurement time $T_m$. Hence we predict that the favorable conditions for the reliable observation of $C_4(\omega)$ should be found at strong beam intensities that increase the probability of light interaction with a spin of a quantum dot.

\section{\label{sec:simple} Stochastic Path Integral Approach Applied to Noninteracting Spin Systems}
Stochastic path integrals are frequently used for the calculation of physical noise statistics.
Here, we will use the version of this technique that was developed by Pilgram {\it et al.} \cite{Pilgram03, Jordan04}, and which is particularly suitable for obtaining cumulants of mesoscopic systems. Originally, this approach was built to study the full counting statistics of electron transport in mesoscopic electric circuits, and then adapted to applications in biochemical reaction networks and explicitly time-dependent systems \cite{Sinitsyn09,Sinitsyn07}.    In this section, we will consider the model of $N$ independent Ising spins as an illustrative example, and in the following two sections we will use this approach to study the effects of ferromagnetic interactions and the Pauli exclusion principle on cumulants of the spin noise.

 The basic idea of constructing the path integral is the separation of time scales. If the number of interacting spins $N$ is mesoscopically large, one can identify the time interval $\Delta t$ such that the number of spins flipped during this time interval is, on average, much larger than unity but still much smaller than $N$. One can consider then the total spin polarization of the observed region as a slow variable, which can be approximated by a constant during time $\Delta t$.
Suppose that the probability $k_1$ is to flip from up to down and the probability $k_2$ is to flip from down to up states of any spin per unit of time. Since the spins do not interact with each other, this model, formally, can be solved without a path integral, as we did in Sec. \ref{sec:single spin} because cumulants for noise of  independent spins simply add. Our goal, however, is to illustrate how path integrals can be used for calculations of higher order spin noise cumulants, in order to apply them later to more complex interacting spin systems.

Let $N_{\ua}$ and $N_{\da}$ be the numbers of observed spins, respectively, in the up and down states.
The average number of spins transferred from up to down during time $\Delta t$ is $\la Q_1 \ra=k_1 N_{\ua} \Delta t$. Similarly, the transferred number of spins from down to up is: $\la Q_2 \ra=k_2 N_{\da}\Delta t$.
$Q_1$ and $Q_2$ are  random variables obeying the Poisson distribution. Their fluctuations result in the variation of the spin polarization, which we define as
$M=N_{\ua}-N_{\da}$.  One can integrate over the fast fluctuating dynamics of $Q_1$ and $Q_2$ to obtain the partition function for the slowly changing variable $M$.
During time $\Delta t$, the variation of $M$ is $M(t+\Delta t)-M(t)=2(Q_2-Q_1)$ and $Q_1$ and $Q_2$ can be considered uncorrelated to each other.

The Poisson probability distributions of $Q_{1,2}$   can be written as  integrals
\be
 P(Q_{1,2})=\int d\chi_{1,2} e^{-i\chi_{1,2} Q_{1,2}+\Delta t H_{1,2}(\chi_{1,2})},
 \label{pq}
 \ee
 in which:
\be
H_1(\chi_1)= k_1 N_{\ua} (e^{i\chi_1}-1),\quad H_2(\chi_2)= k_2 N_{\da} (e^{i\chi_2}-1)
\ee
are the cumulant generating functions of corresponding Poisson distributions.
Following the standard path integral approach  \cite{Pilgram03, Jordan04}, we discretize a long measurement time, $T_m$, into $\Delta t$ segments around time moments: $t_n=n\Delta t$ with $n=1, ..., N$. The sum over all possible system trajectories weighted by their probabilities during this time $T_m$, which is called the  {\it partition function}, can then be written as:
\be
{\cal Z}=\prod_{n} \prod_{j=1, 2} \int dM(t_n) dQ_j(t_n)P(Q_j) \delta \left[M(t_{n+1})-M(t_n)+2(Q_1(t_n)-Q_2(t_n)) \right],
\ee
where the $\delta$-function imposes the conservation constraint. One should then express  the delta function as an integral over a new variable $\chi$, i.e. as $\delta (f)=(2\pi)^{-1}\int d\chi e^{i\chi f}$.
Then one can perform integration over variables $Q_{1, 2}$, and $\chi_{1,2}$, to end up with :
\be
{\cal Z}=\prod_n \int  \frac{dM(t_n) \,d\chi_n}{2\pi} e^{i\chi_n(M(t_{n+1})-M(t_n))+\Delta t (H_1(2\chi_n)+H_2(-2\chi_n))}.
\ee
Taking a continuous limit, we obtain:
\be
{\cal Z}=\int {\cal D} M {\cal D} \chi e^{\int dt (i\chi \dot{M}+H(M, \chi))},
\ee
with the Hamiltonian $H(M, \chi)=H_1(M, 2\chi)+H_2(M, -2\chi)$.

 Even though the Hamiltonian is not a real function of its variables, we can still apply the Hamiltonian formalism to the above path integral.  The variation of the action with respect to $M$ and $\chi$ gives the saddle point solution $\chi=\chi_C$, $M=M_C$, where
\be
i\dot{M}_C=-\frac{\partial H}{\partial \chi_C}, \quad i\dot{\chi}_C=\frac{\partial H}{\partial M_C}.
\label{cl1}
\ee
At the steady state, Eq. (\ref{cl1}) has a solution
\be
\chi_C=0, \quad M_C=\frac{k_2-k_1}{k_1+k_2} N.
\label{semicl}
\ee
One can find that $M_C$ coincides with the mean value of the spin polarization in the system.
In order to obtain higher order spin correlators, one should consider the action near the saddle point (\ref{semicl}) beyond the classical limit and introduce finite values $\chi$ and $\delta M$ that describe the deviation from (\ref{semicl}).
As it was discussed in \cite{Jordan04}, in order to calculate the $n$-th order correlator of variables, it is sufficient to find the Hamiltonian in the action up to the $n$-th order in total powers of $\chi$ and $\delta M$. Hence, we write the partition function as
\be
{\cal Z}= \int {\cal D} M {\cal D} \chi e^{\int dt {\cal L} }, \quad {\cal L}={\cal L}_2+{\cal L}_3+{\cal L}_4+\ldots,
\label{part11}
\ee
where
\be
{\cal L}_2&&=i\chi \delta\dot{ M}+i\gamma\delta M\chi-a\chi^2, \nonumber \\
{\cal L}_3&&=-(k_1-k_2)\delta M\chi^2, \quad  {\cal L}_4=-\frac{2i\gamma}{3}\delta M \chi^3+\frac{a}{3}\chi^4.
\ee
To shorten our notation,  we introduce parameters
\be
\gamma=k_1+k_2,\,\,\,\ {\rm and} \quad a=\frac{4k_1k_2N}{k_1+k_2}.
\label{par1}
\ee
Keeping only the quadratic part of the Lagrangian, ${\cal L}_2$ in (\ref{part11}), one can calculate the second order correlation function by writing the action in the frequency domain. By substituting $\chi(t)=T_m^{-1}\sum_{\omega}e^{i\omega t} \chi(\omega)$ and $\delta M(t)=T_m^{-1}\sum_{\omega}e^{i\omega t} \delta M (\omega)$ into the action, we find
\be
\int dt   {\cal L}_2=\sum_{\omega>0} {\bf a}^*(\omega) \hat{A} {\bf a}(\omega),
\ee
with ${\bf a}=(\chi, \delta M)^T$ and the matrix $\hat{A}$ given by:
\be
\hat{A}=\frac{1}{T_m}\left(\begin{array}{cc}
          -2a & -\omega+i\gamma \\
          \omega+i\gamma & 0
        \end{array} \right).
\ee
Similarly, one can write the higher order contributions to the Lagrangian in the frequency domain, e.g.,
\be
\int dt {\cal L}_3=-(k_1-k_2)\frac{1}{T_m^2}\sum_{\omega_1, \omega_2}\delta M(\omega_1)\chi(\omega_2)\chi(-\omega_1-\omega_2),
\ee
\be
\int dt {\cal L}_4=&&\frac{1}{T_m^3}\sum_{\omega_1, \omega_2, \omega_3} \Big{(} \frac{a}{3}\chi(\omega_1)\chi(\omega_2)\chi(\omega_{3})\chi(-\omega_1-\omega_2-\omega_3) \\
&&-\frac{2i\gamma}{3}\delta M(\omega_1)\chi(\omega_2)\chi(\omega_3)\chi(-\omega_1-\omega_2-\omega_3) \Big{)}.
\ee
The 2nd order correlators are found from (we recall that ${\cal Z}=1$):
\be
\la a_i(-\omega) a_j(\omega)\ra=\int {\cal D} M {\cal D}\chi a_i(-\omega) a_j(\omega) e^{\int dt {\cal L} _2} = \left(- \hat{A}^{-1}\right)_{ij}.
\label{corr21}
\ee
Explicitly:
\be
\la\delta M(-\omega)\delta M(\omega) \ra_0 &&=\frac{2aT_m}{\omega^2+\gamma^2}, \label{eq:C2MM}\\
\la \delta M(-\omega) \chi(\omega) \ra_0 &&=T_m\frac{\omega+i\gamma}{\omega^2+\gamma^2},  \label{eq:C2xM}\\
\la \delta M(\omega) \chi(-\omega) \ra_0 &&=T_m\frac{-\omega +i\gamma}{\omega^2+\gamma^2}  \label{eq:C2Mx}, \quad \la\chi(-\omega)\chi(\omega)\ra_0=0.
\ee
Eq.~(\ref{eq:C2MM}) provides the spin noise power in the frequency domain. As expected, its value coincides with the noise power of the single Ising spin, Eq.~(\ref{eq:C2single}), up to the factor $N$. Correlators of $\delta M$ with $\chi$ do not describe a measurable characteristic but they are needed for the calculation of higher order cumulants.

\begin{figure}
\includegraphics[scale=0.8]{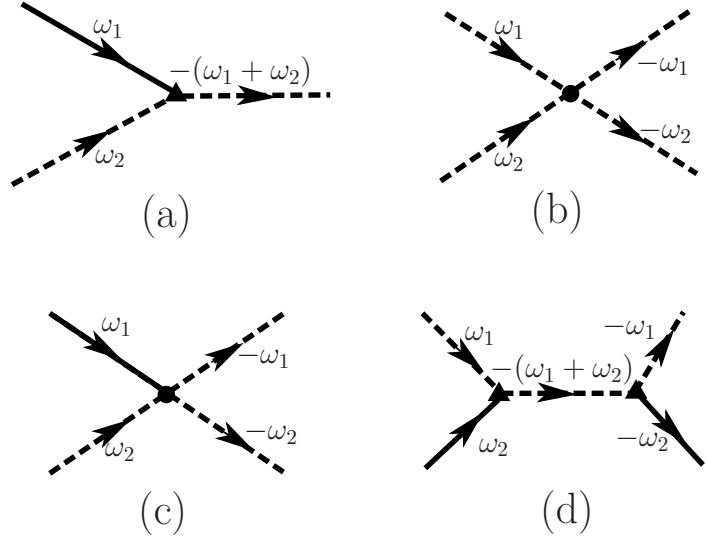}
\caption{\label{fig:diagram}  The Feynman diagrams for $C_3$ (a) and $C_4$ (b, c and d). (a) is the contribution from ${\cal L}_3$. (b) and (c) are the contributions to $C_4(\omega_1, \omega_2)$ due to the two terms of ${\cal L}_4$, respectively. (d) is the tree level contribution to $C_4$ due to ${\cal L}_3$, which is of the  order of (b) and (c) but vanishes when $k_1=k_2$. The solid line denotes $\la \delta M(\omega)\delta M(-\omega) \ra_0$, the dashed line with notation $\omega$ denotes $\la \delta M(\omega) \chi(-\omega) \ra_0$ and the dashed line with $-\omega$ denotes $\la \chi(\omega) \delta M(-\omega) \ra_0$. Other contributing diagrams are obtained by all possible permutations of arrows at each node such that the sum of ingoing and outgoing frequencies is zero.}
\end{figure}

To estimate the  third order cumulant, we keep the term with  ${\cal L}_3$ in the action and treat it as a small perturbation:
\be
C_3\equiv \int {\cal D} M {\cal D} \chi e^{\int dt   {\cal L}_2} \delta M(-\omega_1-\omega_2) \delta M(\omega_1) \delta M(\omega_2) \int dt{\cal L}_3.
\label{quadr-c3}
\ee
Since the exponent in (\ref{quadr-c3}) is quadratic in variables, one can calculate this expression using the Wick rule by summing over all possible products of 2nd order correlators.
Terms in such expressions can be represented by
the Feynman diagrams, e.g.  one of them that contributes  a nonzero value to the cumulant (\ref{quadr-c3}) is shown in Fig. \ref{fig:diagram}(a). Other nonvanishing diagrams contributing to $C_3$ are obtained by all possible permutations of arrows entering the node such that the total sum of ingoing and outgoing frequencies is zero.
 Evaluating these diagrams, we obtain:
\be
C_3=4(k_1-k_2)a T_m \frac{\omega_1^2+\omega_1\omega_2+\omega_2^2+3\gamma^2}{(\omega_1^2+\gamma^2)(\omega_2^2+\gamma^2)[(\omega_1+\omega_2)^2+\gamma^2]}.
\ee
The fourth order cumulant defined in Eq. (\ref{c4-1}), at $k_1=k_2$ can be found as
\be
C_4(\omega_1, \omega_2)=\int {\cal D} M {\cal D} \chi e^{\int dt   {\cal L}_2} |\delta M(\omega_1)|^2 |\delta M(\omega_2)|^2  \int dt {\cal L}_4(t).
\ee
For the time-reversal symmetric case $k_1=k_2 \equiv k$ the result is represented by the sum of diagrams shown in Fig. \ref{fig:diagram} (b) and (c): $C_4=C_4^{(1)}+C_4^{(2)}$, where
\be
C_4^{(1)}&=&8a\frac{T_m}{(\omega_1^2+\gamma^2)(\omega_2^2+\gamma^2)} \label{eq:C41}, \\
C_4^{(2)} &=&-16a\gamma^2 T_m \frac{\omega_1^2+\omega_2^2+2\gamma^2}{(\omega_1^2+\gamma^2)^2(\omega_2^2+\gamma^2)^2}. \label{eq:C42}
\ee
Generally, there is also a tree diagram contribution to $C_4$, which appears when the second power of ${\cal L}_3$ is included in the perturbative expansion, as shown in Fig. \ref{fig:diagram}(d). It is also proportional to the measurement time, $T_m$, but it vanishes at  identical rates of up-down and down-up transitions ($k_1=k_2$), so we do not consider it here.

Finally, we discuss the relative strength of  $C_4$ and $C_2$.  It is useful for this to introduce a dimensionless combination
\be
\eta\equiv\frac{\int d\omega_1 d\omega_2 C_4(\omega_1, \omega_2)}{\int d\omega_1d\omega_2 C_2(\omega_1)C_2(\omega_2)}.
\label{eta}
\ee
At $k_1=k_2=k$, we obtain $\eta=-\frac{2}{aT_m}=-\frac{1}{k NT_m}$.
Notice that when the number of spins is reduced, $C_4$ becomes larger as compared to $C_2(\omega_1)C_2(\omega_2)$. For a single spin, $\eta=\frac{1}{kT_m}$.

Another type of 4th order cumulant,  given in Eq.~(\ref{c4-2}), is characterized by the 4th order to 2nd order cumulant  ratio defined in Eq. \ref{eta-21}. Our analogous calculations give $\eta_1=-\frac{2}{kT_m}=-\frac{2T}{T_m}$ with the characteristic life time $T=1/k$.

\section{\label{sec:ferro} Model with a ferromagnetic coupling}
In order to explore many-body effects on spin noise statistics, here we will study a model with the  Glauber   dynamics of ferromagnetically coupled Ising spins \cite{flindt-13pre}.  We assume that all $N$ observed spins experience an effective magnetic field proportional to the instantaneous spin polarization, $B_z(t) \sim M(t)$, so that the kinetic rates are modified to account for this field and the detailed balance conditions (\ref{k2k1}).
Here, for simplicity, we define $M=\frac{N_{\ua}-N_{\da}}{N}$,  which is slightly different from the quantity defined in the noninteracting case.

In comparison to the noninteracting spin model of the previous section, we assume that $k_1$ and $k_2$ are no longer constant but rather depend on the local magnetization such that $k_1/k_2=e^{-\alpha M}$, where  $\alpha$ is a parameter that characterizes the exchange coupling in the mean field approximation. This parameter also absorbs the inverse temperature $\alpha\propto T_s^{-1}$ in (\ref{k2k1}). We choose $k_1=ke^{-\alpha M/2}$ and $k_2=k e^{\alpha M/2}$, where $k$ is the characteristic kinetic rate at zero $M$. Applying the rules for constructing an effective Hamiltonian in the path integral, we find
\be
H=ke^{-\alpha M/2} \frac{1+M}{2} (e^{i2\chi}-1)+ke^{\alpha M/2} \frac{1-M}{2} (e^{-i2\chi}-1). \label{eq:H2}
\ee
For such a choice of variables $M$ and $\chi$, the partition function is expressed in the form
\be
{\cal Z}= \int {\cal D} M {\cal D} \chi e^{N\int dt (i \chi \dot{M}+H )},
\label{part1}
\ee
in which it is explicitly clear that the correlators, e.g. $\langle M(\omega) M(-\omega) \rangle$, depend as $1/N$ on the total number of observed spins. By recalling that the full polarization is obtained by changing variables $M \rightarrow MN$, one can conclude that all cumulants of the total spin polarization will be finally proportional to $N$.

We will assume that parameter $\alpha$ is tuned so that the system is close to a ferromagnetic phase transition, so that the mean magnetization is either zero or small.
In such an approximation, the saddle point equations for Eq.~(\ref{eq:H2}) have two solutions: one is $\chi=0$ and $M=0$; the other is $\chi=0$ and $M^2\approx \frac{4}{\alpha^2}(\frac{\alpha -2}{1-\alpha /6})$. This means that there is a critical value of the parameter $\alpha$, namely, $\alpha_c=2$ that corresponds to the phase transition between paramagnetic and ferromagnetic phases.

\subsection{Correlations in the paramagnetic phase (temperatures above the phase transition)}
\begin{figure}
\includegraphics[scale=0.42]{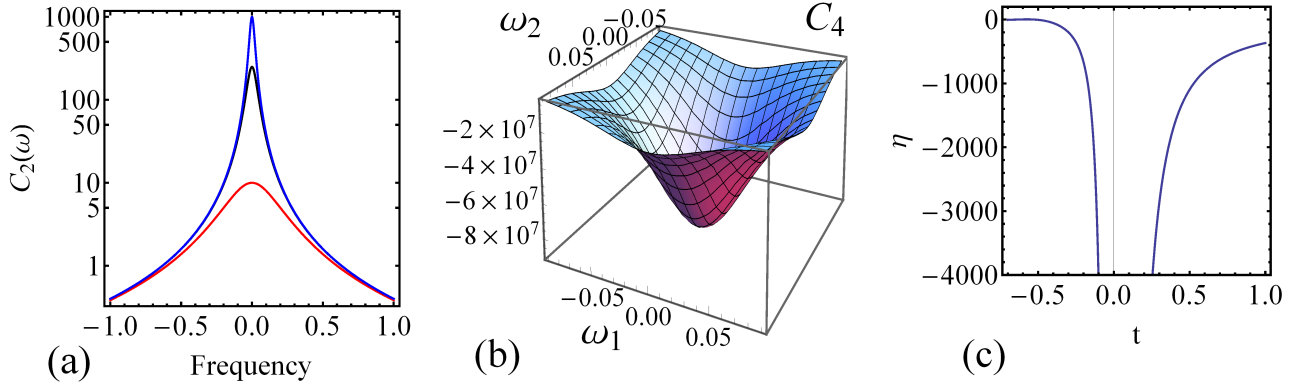}
\caption{\label{fig:C2C4eta}  (a) The second order cumulant  for a single spin with $k_1=k_2=0.1$ (red) and for a spin system with ferromagnetic interaction with $k=0.1$ and $t=0.1$ (blue) and $t=0.2$ (black). (b) Fourth order cumulant for a spin system with ferromagnetic interaction in the paramagnetic phase. $k=0.2$ and $t=0.2$. (c) The quantity $\eta$ defined in the text which shows the relative strength of $C_4$ and $C_2$ as a function of normalized temperature $t=(T-T_c)/T_c$.}
\end{figure}

At temperatures slightly above the phase transition point, we would have zero average magnetization, $M=0$.
We introduce a small parameter, $t= (T_s-T_C)/T_C$, such that $\alpha \approx 2(1-t)$,  and expand the Lagrangian up to  $4$th order in powers of fluctuations from this point: ${\cal L}={\cal L}_2+{\cal L}_3+{\cal L}_4+\ldots$:
\be
{\cal L}_2&&=i\chi \delta\dot{ M}+i \gamma \delta M \chi- a\chi^2, ~~~~{\cal L}_3=0,  \nonumber \\
{\cal L}_4&&=\frac{ia}{3}\delta M^3\chi+\frac{a}{2}\delta M^2\chi^2 - \frac{i2\gamma}{3} \delta M\chi^3+\frac{a}{3}\chi^4,
\label{l4f}
\ee
where
\be
 a=2k, \quad \gamma=2kt.
 \label{parf}
 \ee
Evidently, due to the time-reversal symmetry in the paramagnetic phase, the third order  cumulant is zero.  The second order cumulant has the same Lorentzian form as noninteracting spins in Eq. (\ref{eq:C2MM}), (\ref{eq:C2xM}), and (\ref{eq:C2Mx}), but with parameters $a$ and $\gamma$ defined in (\ref{parf}).
At the phase transition, the effective relaxation rate vanishes $\gamma\rar 0$, indicating the critical slowdown. Hence, the amplitude of the 2nd order correlator  grows as it approaches the phase transition, as shown in Fig.~\ref{fig:C2C4eta}(a).

Each term in ${\cal L}_4$  corresponds to some contribution to the 4th cumulant of spin noise, so that
\be
C_4(\omega_1,\omega_2) = C_4^{(1)}+C_4^{(2)}+C_4^{(3)}+C_4^{(4)}.
\label{c4123}
\ee
The first two terms in (\ref{c4123}) are due to $\chi^4$ and $\delta M\chi^3$ in (\ref{l4f}). We have previously calculated them for the independent Ising spins model. They produce the same expressions as Eq. (\ref{eq:C41}) and Eq. (\ref{eq:C42}) with  redefined parameters according to (\ref{parf}).
The terms $\sim \delta M^2\chi^2$ and $\sim \delta M^3\chi$ in (\ref{l4f}) are new. Their contributions to $C_4$ are given by
\be
C_4^{(3)}&&=-8a^3 NT_m\frac{\omega_1^2+\omega_2^2+6\gamma^2}{(\omega_1^2+\gamma^2)^2(\omega_2^2+\gamma^2)^2}, \\
C_4^{(4)}&&=-64a^4 NT_m \frac{\gamma}{(\omega_1^2+\gamma^2)^2(\omega_2^2+\gamma^2)^2}.
\ee
In Fig.~\ref{fig:C2C4eta}(b), we plot $C_4(\omega_1, \omega_2)$ in the paramagnetic phase, according to which, the 4th order correlator is still negative but it can have a very large amplitude near the phase transition.

\subsection{Spin noise in the ferromagnetic state (temperature below phase transition)}
A specific feature of the ferromagnetic state is the absence of the time-reversal symmetry, meaning that the 3rd order cumulant of the spin noise may become nonzero.
Below the critical temperature,  we would have the saddle point, which corresponds to $M=\pm M_0=\pm \sqrt{3 |t| }$. Expanding the Lagrangian up to the 3rd power of fluctuations near $M=M_0$, we find
\be
{\cal L}=i\chi \delta\dot{ M}-2k\chi^2+4ik |t| \chi \delta M
+aM_0\chi^2\delta M+iaM_0 \chi\delta M^2. 
\label{l33}
\ee
In this case, $a=2k$, $\gamma=4k|t|$. Note that the relaxation rate is twice the one in the paramagnetic phase at the same distance to the phase transition point.
Diagrammatic calculations show that the two 3rd order terms in ({\ref{l33}) produce two contributions to the 3rd order cumulant:
\be
C_3=C_3^{(1)}+C_3^{(2)},
\ee
where
\be
C_3^{(1)}&&=-4a^2NM_0 T_m\frac{\omega_1^2+\omega_2^2+\omega_1\omega_2+3\gamma^2}{(\omega_1^2+\gamma^2)(\omega_2^2+\gamma^2)[(\omega_1+\omega_2)^2+\gamma^2]},\\
C_3^{(2)}&&=-24a^3NM_0 T_m\frac{\gamma}{(\omega_1^2+\gamma^2)(\omega_2^2+\gamma^2)[(\omega_1+\omega_2)^2+\gamma^2]}.
\ee
Both terms depend linearly on the average magnetization $M_0$, which in turn scales as $M_0 \sim |t|^{1/2}$ as a function of temperature distance to the phase transition.

The 4th order cumulant in this phase would be very complicated to show here properly. In Fig.~\ref{fig:C2C4eta}(c) we  just plot the numerical result for  the dimensionless parameter $\eta$ as a function of $t$ for both $t>0$ and $t<0$. Since the parameter $\eta$ characterizes the deviation of the distribution of the spin noise from a Gaussian form, its divergence at the phase transition point indicates a strongly non-Gaussian spin noise statistics. A  difference between $t>0$ and $t<0$ could be  traced to the difference of relaxation rates $\gamma$ in the two cases.

\section{\label{sec;Pauli}Conducting electrons with Pauli exclusion principle}
As another example of nontrivial higher order correlations, consider the Fermi sea of conducting electrons.  Previously charge current fluctuations in such systems have been studied extensively. Partial suppression of the shot noise by the Pauli principle has been one of the most  important effects in this field.  It was studied previously, in particular, by the method of the stochastic path integral \cite{Jordan04}. Here we will explore effects of the Pauli exclusion principle on the cumulants of the local spin noise fluctuations.
We assume that, due to the phonon coupling in the observation region, the local electron distribution in the momentum space, for each spin species, quickly equilibrates and restores to the Fermi-Dirac distribution at the ambient temperature $T_s$ and at local chemical potentials $\mu_{\ua} ( t)$ and $\mu_{\da}(t)$ for, respectively, spin up and spin down electrons. We assume that the spin degree of freedom equilibrates at a much longer time scale, e.g. the spin relaxation due to the Dyakonov-Perel  mechanism is at $\sim 100$ns, while thermalization of orbital degrees of freedom can be at sub-nanoseconds for conducting electrons in GaAs. Due to spin flipping, chemical potentials, $\mu_{\ua} (t)$ and $\mu_{\da}(t)$  will fluctuate.
We normalize the chemical potentials so that at a zero net spin polarization they are set to zero. The excess of electrons with spin up then can be related to the chemical potential by

\be
N_{\ua} = D\int d\epsilon  \left( \frac{1}{1+e^{(\epsilon-\mu_{\ua})/T_s}} - \frac{1}{1+e^{\epsilon/T_s}} \right)=D\mu_{\ua},
\label{constr}
\ee
where $D$ is the density of states in the observation region per spin and per unit of energy near the Fermi surface. Note that $D$ is not an intensive characteristic in the sense that it is not a density per volume of the system. For example  it is proportional to the size of the mesoscopic observation region in Fig.~1.
We set the Boltzmann constant $k_B=1$. For a sufficiently large observation region, electroneutrality ensures that $N_{\ua}=-N_{\da}$.

\begin{figure}
\includegraphics[scale=0.5]{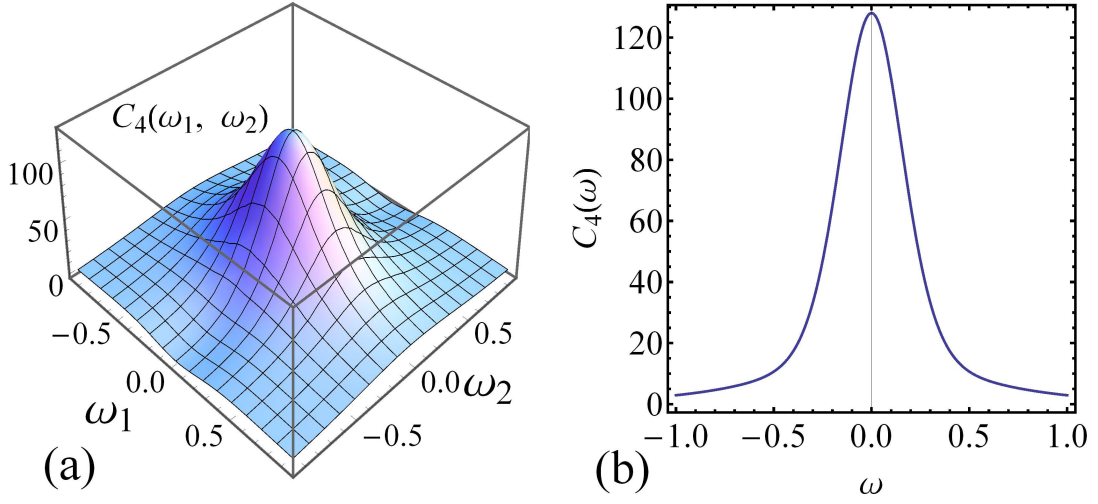}
\caption{\label{c4F}  Fourth cumulants, (a) $C_4(\omega_1,\omega_2)$ and (b) $C_4(\omega)$, in the model of spin fluctuations with Pauli exclusion interactions. Parameters: $ a=1$ and $\gamma=0.5$.}
\end{figure}

To account for  the Pauli exclusion principle, we assume that the average number of spin flips, e.g. from up to down, for  electrons with energy $\epsilon$ is proportional to the number of electrons in the observation region with spin up and the density of unfilled states with the spin down at energy $\epsilon$.
Then the total average rate of transitions from up to down is given by:
\be
J_{\ua  \da}&=&k D\int d\epsilon \frac{1}{1+e^{(\epsilon-\mu_{\ua})/T_s}}\Big{(} 1-\frac{1}{1+e^{(\epsilon-\mu_{\da})/T_s}} \Big{)}
=k \frac{D(\mu_{\ua}-\mu_{\da})}{1-e^{(\mu_{\da}-\mu_{\ua})/T_s}},
\ee
where $k$ is the characteristic transition rate.
Similarly,
\be
J_{\da \ua}=k \frac{D(\mu_{\da}-\mu_{\ua})}{1-e^{(\mu_{\ua}-\mu_{\da})/T_s}}.
\ee

The local spin polarization is $M=N_{\ua}-N_{\da}$. Since a single spin flip changes the spin polarization by 2,
 the Hamiltonian in the path integral action that describes the dynamics of $M$ is written as:
\be
H=J_{\ua  \da}(e^{2i\chi}-1)+J_{\da \ua}(e^{-2i\chi}-1),
\ee
where $\chi$ is the variable conjugated to $M$. Explicitly:
\be
H=\frac{k M}{1-e^{- M/(DT_s)}}(e^{i2\chi}-1)+\frac{-k M}{1-e^{M/(DT_s)}}(e^{-i2\chi}-1).
\ee
The saddle point solution corresponds to $\chi_C=0$ and $M_C=0$. Expanding the Lagrangian up to the fourth order, we obtain:
\be
{\cal L}=&&i\chi \delta\dot{ M}+2ik  \delta M\chi-4k DT_s\chi^2
-\frac{k}{3DT_s}\delta M^2\chi^2-\frac{4ik}{3} \delta M\chi^3+\frac{4kDT_s}{3}\chi^4.
\label{lag-fermi}
\ee
The second order cumulant would have the same Lorentzian shape as it would be for independent Ising spins with parameters $a=4k DT_s$, $\gamma=2k $. The Pauli exclusion principle changes the expression of $a$ by replacing the total number of spins $N$ with the number of spins near the Fermi surface in the interval of energy of the order of temperature $T_s$.
The fourth order cumulant is the sum of three terms that correspond to contributions of each of the three terms in ${\cal L}_4$:
\be
C_4=C_4^{(1)}+C_4^{(2)}+C_4^{(3)},
\label{c4-fermi}
\ee
where
\be
C_4^{(1)}&&=8a\frac{T_m}{(\omega_1^2+\gamma^2)(\omega_2^2+\gamma^2)}, \\
\label{c411}
\\
C_4^{(2)}&&=-16a\gamma^2 T_m \frac{\omega_1^2+\omega_2^2+2\gamma^2}{(\omega_1^2+\gamma^2)^2(\omega_2^2+\gamma^2)^2},
\label{c412}
\ee
\be
C_4^{(3)}&&=\frac{16a\gamma^2T_m}{3}\frac{\omega_1^2+\omega_2^2+6\gamma^2}{(\omega_1^2+\gamma^2)^2(\omega_2^2+\gamma^2)^2}.
\label{c413}
\ee
The first two terms have the same form as for  noninteracting Ising spins, (one can compare (\ref{c411}), (\ref{c412}) with Eq. (\ref{eq:C41}) and Eq. (\ref{eq:C42})).
The new contribution (\ref{c413}) appears from the term $\sim\delta M^2\chi^2$ in (\ref{lag-fermi}).
It radically changes properties of the bispectrum, as shown in Fig.~\ref{c4F}. In particular,
the dimensionless parameter $\eta$ for the Fermi system becomes positive, $\eta=\frac{2}{3a}=\frac{1}{6k DT_s T_m}$. Comparing this result with noninteracting and ferromagnetic interaction cases, for which $\eta$ is negative,  we conclude that the functional form of the fourth order cumulant is sensitive to the details of  spin interactions. Note also that at lower temperatures, the Pauli exclusion principle makes $\eta$ larger, i.e. statistics becomes less Gaussian.

\section{Discussion}

Higher order cumulants, by construction, contain additional information to the noise power spectrum. For example, in all models considered in this work, the noise power has a Lorentzian form, from which only a single parameter, i.e. the effective relaxation time can be determined. In contrast, the 3rd order cumulant contains information about the asymmetry of relaxation rates, while measurements of the 4th order cumulant can be used to estimate not only all parameters of the considered models but also distinguish among candidate models if the Hamiltonian of the spin system is not a priory known, because the functional form and the sign of the 4th order cumulant are sensitive to subtle details of the  kinetics. We also predict that  ratios of cumulants, such as parameters $\eta_1$ and $\eta$ in Eqs.~(\ref{eta-21}) and (\ref{eta}) provide a good estimate of the size of physical spin correlations.

We explored the properties of higher order cumulants of the spin noise in the frequency domain and discussed the conditions for their experimental observation. In a standard framework of most of the SNS experiments, measurements were performed on a mesocopic number, e.g. $N\sim 10^5$, of independent spins. In such a case, the higher cumulants are suppressed in comparison to the noise power, e.g. $C_4/(C_2)^2 \sim 1/N$. Our results suggest two strategies that can be used to enhance this ratio. First, one can perform measurements on
a smaller number of spins. Since the spin noise characterization of a single spin is now accessible in InGaAs hole doped quantum dots \cite{Dahbashi13}, we believe the measurements of higher order noise cumulants in such systems are already possible. Our analytical and numerical calculations predict a negative value of the 4th order cumulant at a zero external magnetic field. The 3rd order cumulant can be also observed in such systems in a strong out-of-plane magnetic field.

The second strategy to observe higher order spin cumulants is to perform measurements on strongly interacting spins. In the case of ferromagnetic interactions, spin fluctuations are strongly enhanced near the paramagnetic/ferromagnetic phase transition, leading to a much stronger signal for all cumulants. In such a case, spins flip not independently but rather as clusters of many correlated spins, and the number  $N$ should be interpreted as the typical number of clusters in the observation region, which can be considerably smaller than the total number of observed spins. Hence, we predict that the higher order  cumulants can be substantially more important for the characterization of magnetic semiconductors, especially near the  phase transition temperatures.

To study spin noise in interacting spin systems, we  developed a quantitative theoretical approach, which is based on the stochastic path integral technique. We calculated higher order cumulants in a model of a ferromagnetically coupled interacting spin system.
At temperatures below the phase transition point, we found that the 3rd spin cumulant becomes nonzero in the state with a spontaneous  symmetry breaking. Approaching the phase transition point from higher temperatures, the higher order cumulants grow not only in the magnitude but also in comparison to the noise power.

 Finally, by applying the stochastic path integral technique to conducting electrons we found an  enhancement of the relative role of $C_4$ due to the Pauli exclusion principle. In the metallic phase, we predict that the dimensionless ratio of $C_4$ and $(C_2)^2$ is inversely proportional to temperature. Hence, by observing the spin noise at the lowest possible temperatures one may achieve the regime with strongly non-Gaussian spin fluctuations.

\section*{Acknowledgement.}

We thank Yan Li and S. A. Crooker for useful discussions. Work at LANL was carried out under the auspices of the Project No. LDRD/20110189ER and the National Nuclear Security Administration of the U.S. Department of Energy at Los Alamos National Laboratory under Contract No. DE-AC52-06NA25396.

\section*{References}
\bibliography{aipsamp}


\end{document}